\documentclass[final,3p,times,twocolumn]{elsarticle}

\usepackage[utf8]{inputenc}
\usepackage{amsmath,amssymb,exscale}
\usepackage{graphicx}
\usepackage{epsfig}
\usepackage{multicol}
\usepackage{color}
\usepackage{xcolor}
\usepackage{hyperref}
\usepackage{mathrsfs}
\hypersetup{colorlinks,bookmarksopen,bookmarksnumbered,citecolor=rossos,
linkcolor=black,pdfstartview=FitH,urlcolor=blus}
\usepackage{amsfonts}
\usepackage{slashed}
\usepackage{blindtext}
 \usepackage{fancyhdr}
\usepackage{mathtools}
\usepackage{cancel,xcolor}
\biboptions{numbers,sort&compress}
\usepackage{tensor}
\usepackage{soul}

\allowdisplaybreaks

\newcommand{\be}{\begin{equation}}
\newcommand{\ee}{\end{equation}}
\newcommand{\bear}{\begin{array}}
\newcommand{\eear}{\end{array}}
\newcommand{\ba}{\begin{eqnarray}}
\newcommand{\ea}{\end{eqnarray}}
 \def\a{\alpha}
\def\b{\beta}

\def\g{\gamma}
\def\k{\kappa}
\def\l{\lambda}
\def\m{\mu}
\def\n{\nu}

\def\r{\rho}
\def\s{\sigma}

\def\x{\xi}

\def\tns{\tensor}
 
\definecolor{blue(pigment)}{rgb}{0.2, 0.2, 0.6}
\definecolor{coolblack}{rgb}{0.0, 0.18, 0.39}
\definecolor{darkblue}{rgb}{0.0, 0.0, 0.55}
\definecolor{darkcerulean}{rgb}{0.03, 0.27, 0.49}
\definecolor{violetred}{rgb}{0.8,0.13,0.56}
\definecolor{fuxia}{rgb}{1,0,1}
\definecolor{orchid}{rgb}{0.85,0.44,0.84}
\definecolor{lightpink}{rgb}{1.00,0.71,0.76}

\let\oldmaketitle\maketitle
\renewcommand{\maketitle}{\oldmaketitle\setcounter{footnote}{0}}

\begin{document}

\begin{frontmatter}

\title{\vspace{-1cm}\huge {\color{blue(pigment)}Bimetric-Affine Quadratic Gravity} \vspace{0.5cm}
}  
 \author[inst1]{\large {\bf Ioannis D. Gialamas}\corref{cor1}}
 \cortext[cor1]{Email address: ioannis.gialamas@kbfi.ee}
\address[inst1]{\normalsize Laboratory of High Energy and Computational Physics, 
National Institute of Chemical Physics and Biophysics, \\R{\"a}vala pst.~10, 10143, Tallinn, Estonia}

\author[inst2]{\large {\bf Kyriakos Tamvakis}\fnref{cor2}}
\fntext[cor2]{Email address: tamvakis@uoi.gr}
\address[inst2]{\normalsize Physics Department, University of Ioannina, 45110, Ioannina, Greece}

\begin{abstract}
Bimetric gravity is a theory of gravity that posits the existence of two interacting and dynamical metric tensors. The spectrum of bimetric gravity consists of a massless and a massive spin-2 particle. The form of the interactions between the two metrics $g_{\m\n}$ and $f_{\m\n}$ is constrained by requiring absence of the so called {\textit{Boulware-Deser}} ghost. In this work we extend the original bimetric theory to its {\textit{bimetric-affine}} counterpart, in which the two connections, associated with the Ricci scalars,  are treated as independent variables. We examine in detail the case of an additional quadratic in the Ricci scalar curvature term $\mathcal{R}^2(g,\Gamma)$ and we find that this theory is free of ghosts for a wide range of the interaction parameters, not excluding the possibility of a dark matter interpretation of the massive spin-2 particle.
\end{abstract}

\end{frontmatter}
 
 \begingroup
\hypersetup{linkcolor=blus}
\endgroup

\section{Introduction}\label{introduction}
\vspace{-0.1cm}
The framework of modern cosmology, consisting of the theory of gravitation and the Standard Model of particle physics is characterized by the fact that the former is treated classically due to the present lack of a fully quantum UV completion of general relativity (GR), in contrast to the latter which is a fully quantum field theory. Nevertheless, as long as we are not very close to the Planck energy scale it is believed that the effects of gravitation can be treated classically in terms of an effective theory of gravity based on GR with possible modifications arising from the quantum nature of gravitating matter fields. Independently of such expected modifications, open cosmological issues like the dark energy associated with the present accelerated expansion might require the modification of GR at large distances as well. The problem of dark energy, the equally pressing problem of dark matter as well as the realization of inflation at the early stages of the Universe have attracted a number of proposals based on the introduction of new fields, mostly scalar but also fermions and vectors. New fields directly related to the gravitational sector such as the scalar mode arising in the Starobinsky model~\cite{Starobinsky1980} or a massive spin-$2$ partner of the graviton have also been considered, the latter being the key ingredient of the bimetric theory of gravity~\cite{Hassan:2011zd}. A massive graviton also arises in the massive gravity theory of Dvali–Gabadadze–Porrati~\cite{Dvali:2000hr, Dvali:2000xg}. The theory of a massive spin-$2$ field has been first considered by Fierz and Pauli~\cite{Fierz:1939ix}. Although the associated problem of a smooth massless limit was resolved~\cite{Vainshtein:1972sx}, Boulware and Deser~\cite{Boulware:1972yco} showed that as it stood the theory would necessarily contain a ghost. De Rham, Gabadadze, and Tolley (dRGT) studied a nonlinear theory of massive gravity in whose decoupling limit they proved the absence of the ghost degree of freedom~\cite{deRham:2010kj}, while Hassan and Rosen~\cite{Hassan:2011vm} generalized the dRGT massive gravity by including an arbitrary  reference metric $f_{\m\n}$ instead of the Minkowski one (see also~\cite{Hassan:2011tf, Hassan:2011hr, Hassan:2012qv, Hassan:2011ea}). At relatively recent time a ghost-free nonlinear theory of massive spin-$2$ was formulated by Hassan and Rosen~\cite{Hassan:2011zd}, but this time the gravitational action includes an additional Ricci scalar for the new metric $f_{\m\n}$. A key feature of the new theory, named bimetric gravity (or bigravity), is that both metrics are dynamical contrary to the massive gravity theories so far. For reviews on the subject of massive gravity and bimetric gravity see \textit{e.g.}~\cite{Hinterbichler:2011tt, deRham:2014zqa, Schmidt-May:2015vnx}.

Bimetric gravity has been studied extensively in recent years with still continuing activity~\cite{Hassan:2012wr, Hassan:2012gz, Hassan:2013pca, Gording:2018not,  Lust:2021jps}. An aspect of bimetric gravity of a phenomenological interest is the possible interpretation of the massive spin-2 particle as a dark matter candidate~\cite{Aoki:2014cla, Aoki:2016zgp, Babichev:2016hir, Babichev:2016bxi, Marzola:2017lbt, GonzalezAlbornoz:2017gbh, Manita:2022tkl, Kolb:2023dzp}. In this scenario the coupling of the massive spin-2 particle to the Standard Model particles, being of gravitational origin, is naturally Planck suppressed. This exceptionally feeble coupling to matter could clarify why dedicated detection experiments and collider searches have not found any signals of dark matter.

As it is mentioned above gravitating quantum matter fields are bound to generate modifications of GR, a well-known example being the Starobinsky model~\cite{Starobinsky1980} featuring an additional quadratic Ricci scalar curvature $R^2$ term. An immediate question arising is whether the ghost-free construction of bimetric gravity could still be applied appropriately modified. Since the standard metric $R^2$ theory gives rise to an additional dynamical scalar degree of freedom the answer to this question rests on the detailed nonlinear dynamics involving this extra scalar (see~\cite{Nojiri:2012zu, Nojiri:2012re, Kluson:2013yaa, Bamba:2013fha, Nojiri:2015qyc} for applications of general $F(R)$ bimetric extensions under the metric formulation of gravity). In contrast, in the metric-affine (Palatini) formulation~\cite{Sotiriou:2006qn} of the Starobinsky model, in which the connection is an independent variable, no such extra dynamical scalar arises. In such a framework the construction of a ghost-free bimetric $\mathcal{R}^2$ theory, involving only the graviton and its massive spin-$2$ partner, can be conclusive. We note that the presence of quadratic $\mathcal{R}^2$ terms have proven to be important in models of cosmological inflation~\cite{Meng:2004yf, Enckell:2018hmo, Antoniadis:2018ywb, Antoniadis:2018yfq, Gialamas:2019nly, Tenkanen:2020cvw, Lloyd-Stubbs:2020pvx, Antoniadis:2020dfq, Das:2020kff, Gialamas:2020snr, Karam:2021sno, Lykkas:2021vax, Gialamas:2021enw, Antoniadis:2021axu, Lahanas:2022mng, Gialamas:2022xtt}.

In the present article we extend the original bimetric theory to its {\textit{bimetric-affine}} counterpart and study the case of an additional quadratic curvature term, focusing on the case of a quadratic term of the scalar curvature associated to the standard graviton metric. We demonstrate that models of ghost-free\footnote{ See~\cite{BeltranJimenez:2019acz, BeltranJimenez:2020sqf} for a discussion on the presence of ghosts in general metric-affine theories with higher order curvature terms.} gravitating massive spin-$2$ can be constructed along the lines of the standard bimetric gravity theory. In Section~\ref{BAG} we set up the theoretical framework of bimetric gravity and the \textit{bimetric-affine} extension of it. In section~\ref{Q_BAG} we consider \textit{bimetric-affine} theories based on an action that includes quadratic terms of the Ricci scalar $\mathcal{R}^2$ and derive the equivalent bimetric action with a modified potential. The linearized \textit{bimetric-affine} quadratic action is presented in section~\ref{LIN} and in section~\ref{Parameter} we analyze the corresponding parametric space. Finally, we summarize and conclude in section~\ref{Conclusions}.


\vspace{-0.1cm}
 \section{Bimetric Gravity}\label{BAG}
\vspace{-0.1cm}
Standard bimetric gravity~\cite{Hassan:2011zd} based on the Einstein-Hilbert action\footnote{Throughout this paper we use different symbols for the curvature scalar or/and tensors, which in the standard metric gravity we denote by $R$, while in metric-affine gravity by $\mathcal{R}$.} is defined as\footnote{In the literature it is common to replace the scale $m_f$ by a dimensionless parameter $\alpha$ defined through $m_f=\alpha\,m_g$. Also, the scale $m$ is redundant, being an overall scale for the parameters $\beta_n$ and it is often set as  $m^2=\alpha^2 m_g^2$.}
\begin{align}
\mathcal{S}=\int{\rm d}^4x\,\Big\{& m_g^2\sqrt{-g}R(g)+m_f^2\sqrt{-f}R(f) \nonumber
\\& +2m_g^2{m^2}\sqrt{-g}V(\sqrt{\Delta})\Big\}\,,{\label{ACT-0}}
\end{align}
where $g_{ \mu\nu}$ and $f_{ \mu\nu}$ stand for the two metric tensors and the potential depends on the tensor $\left(\sqrt{\Delta}\right)^{ \mu}_{\,\,\,\nu}$, defined by
\be \left(\sqrt{\Delta}\right)^{ \mu}_{\,\,\,\rho}\left(\sqrt{\Delta}\right)^{ \rho}_{\,\,\,\nu}\,=\,\Delta^{ \mu}_{\,\,\,\nu}\,=\,g^{ \mu\rho}f_{ \rho\nu}\,.{\label{DELTA}}\ee
The form of the potential is constrained so that the so-called {\textit{Boulware-Deser}} ghost~\cite{Boulware:1972yco} is absent and the arising mass for one of the spin-2 combinations has the Fierz-Pauli form. The exact form of the potential is
\be 
\label{eq:potV}
V\left(\sqrt{\Delta}\right)\,=\,\sum_{n=0}^4\beta_n\,e_n\left(\sqrt{\Delta}\right)\,,\ee
where $\beta_n$ are parameters and $e_n\left(\sqrt{\Delta}\right)$ are five functions of the metric tensors, which are given by
\be 
e_n\left(\sqrt{\Delta}\right)=\frac{(-1)^{n+1}}{n}\sum_{k=0}^{n-1}(-1)^k {\rm Tr}\left(\sqrt{\Delta}^{n-k}\right)e_k\left(\sqrt{\Delta}\right)\,, \label{eq:polyn}
\ee 
starting with $e_0\left(\sqrt{\Delta}\right)=1$.
To obtain a linearized approximation of the above theory we consider small fluctuations of the metrics around a common background $\bar{g}_{ \mu\nu}$
\be g_{ \mu\nu}\,\approx\,\bar{g}_{ \mu\nu}\,+\,{h_{ \mu\nu}},\,\,\,\,\,f_{ \mu\nu}\,\approx\,\bar{g}_{ \mu\nu}\,+\,{l_{ \mu\nu}}\,.\ee
The resulting linearized action takes the form   
\begin{align}
\label{eq:lin_act}
\mathcal{S} = \int {\rm d}^4x & \sqrt{-\bar{g}} \bigg\{\frac{m_g^2}{4}h_{ \mu\nu}\tns{\mathcal{\mathcal{E}}}{^\m^\n_\r_\s}h^{ \rho\sigma}\,+\,\frac{m_f^2}{4}l_{ \mu\nu}\tns{\mathcal{\mathcal{E}}}{^\m^\n_\r_\s}l^{ \rho\sigma} \nonumber
\\&  +B \left[ \left( \tns{h}{_\m_\n}-\tns{l}{_\m_\n}\right)^2 + C\left( h-l\right)^2 \right]  \nonumber
\\& +  A_g \left( h_{\m\n} h^{\m\n} -\frac{1}{2}h^2 -2h -4\right)  \nonumber
\\&+  A_f \left( l_{\m\n} l^{\m\n} -\frac{1}{2}l^2 -2l-4 \right)  \bigg\}\,, 
\end{align}
where $\tns{\mathcal{\mathcal{E}}}{^\m^\n_\r_\s}$ is the  Lichnerowicz operator.
The appearing parametric coefficients are
\begin{align}
B &=\frac{1}{4} m_g^2 {m^2} \left(\b_1+2\b_2+\b_3 \right)\,, \nonumber
\\A_g &= -\frac{1}{2} m_g^2 {m^2}\left(\b_0 +3\b_1+3\b_2+\b_3 \right)\,, \nonumber
\\A_f &= -\frac{1}{2} m_g^2 {m^2} \left(\b_1+3\b_2+3\b_3+\b_4\right)\,.
\end{align}
 The linearized action~\eqref{eq:lin_act} describes one massless and one massive spin-$2$ particle with zero cosmological constant if
\be
\label{eq:constraints}
A_g=A_f=0\,,\quad C=-1 \quad \text{and} \quad B<0\,.
\ee
Unlike the quadratic theory, which will be analyzed in the next section, in the case of the usual bimetric theory the requirement $C=-1$ that ``removes" the ghost degree of freedom from the theory is automatically satisfied due to the special form of the potential~\eqref{eq:potV}. Note also that in general one parameter is fixed from the equality of the cosmological constants and a second one if we assume zero $A$'s, i.e. zero cosmological constants.

Applying the conditions~\eqref{eq:constraints} we can rewrite the action~\eqref{eq:lin_act} in terms of the massive $(M_{\m\n})$ and  massless $(G_{\m\n})$ eigenstates
\begin{subequations}
\begin{align}
M_{\m\n}&= \frac{m_g m_f}{\sqrt{m_g^2+m_f^2}} \left(l_{\m\n}-h_{\m\n} \right)\,,  \\G_{\m\n}& = \frac{1}{\sqrt{m_g^2+m_f^2}} \left(m_g^2h_{\m\n}+m_f^2 l_{\m\n}\right) \,.
\end{align}
\end{subequations}
The inverse relations are
\begin{subequations}
\begin{align}
h_{ \mu\nu}&=\frac{1}{m_g\sqrt{m_g^2+m_f^2}}\left(m_g G_{ \mu\nu}-m_f M_{ \mu\nu}\right)\,, \\
l_{ \mu\nu}& =\frac{1}{m_f\sqrt{m_g^2+m_f^2}}\left(m_fG_{ \mu\nu}+m_gM_{ \mu\nu}\right)\,.
\end{align}
\end{subequations}
The action is\footnote{Assuming that the coupling to matter is
$$g^{ \mu\nu}T_{ \mu\nu}\,=\,\frac{1}{\sqrt{m_g^2+m_f^2}}G^{ \mu\nu}T_{ \mu\nu}-\frac{m_f}{m_g\sqrt{m_g^2+m_f^2}}M^{ \mu\nu}T_{ \mu\nu}\,,$$
from the effective coupling of the massless mode we can read off the physical Planck mass to be $M_P=\sqrt{m_g^2+m_f^2}$. The issue of which metric tensor should couple to the Standard Model particle spectrum has been discussed extensively in the literature~\cite{Tamanini:2013xia, Aoki:2013joa, Akrami:2014lja, Yamashita:2014fga, deRham:2014naa, Noller:2014sta, Enander:2014xga, Schmidt-May:2014xla, deRham:2014fha, Solomon:2014iwa, Comelli:2015pua, Gumrukcuoglu:2015nua, Blanchet:2015sra, Blanchet:2015bia, Lagos:2015sya, Melville:2015dba, Luben:2018kll}, and as a result, only two feasible options are free of ghost degrees of freedom. One option is for a matter field to couple minimally to only one of the metric tensors. The other option is for matter to couple minimally to an effective metric that is formed by combining the two metrics.}
\begin{align}
\mathcal{S}  =\int  {\rm d}^4x \bigg[ &\frac14 G_{\m\n} \tns{\mathcal{\mathcal{E}}}{^\m^\n_\r_\s} G^{\r\s} +\frac14 M_{\m\n} \tns{\mathcal{\mathcal{E}}}{^\m^\n_\r_\s} M^{\r\s} \nonumber
\\&-\frac{m_{\rm FP}^2}{4} \left(M_{\m\n}M^{\m\n} -M^2 \right) \bigg]\,.
\label{eq:mass_Action}
\end{align}
Note that the Fierz-Pauli mass is
\be m_{\rm FP}^2=-4\frac{m_g^2+m_f^2}{m_g^2m_f^2}B=-\frac{{m^2}(m_g^2+m_f^2)}{{m_f^2}}(\beta_1+2\beta_2+\beta_3)\,.\ee
The so-called {\textit{minimal choice}} for the $\beta$ parameters is 
\be \beta_0=3,\,\,\,\,\beta_1=-1,\,\,\,\,\beta_2=\beta_3=0,\,\,\,\,\beta_4=1\,.\ee
These values yield
\be A_g=A_f=0\,\,\,\,\,\,{\text{and}}\,\,\,\,B=-\frac{1}{4}m_g^2{m^2}\,.\ee 
Therefore, in this case $m_{\rm FP}^2= {m^2}(m_g^2+m_f^2)/{m_f^2}$.

The bimetric action~\eqref{ACT-0} can also be considered in a \textit{bimetric-affine} framework in which the connections associated with each Ricci scalar are not constrained by the Levi-Civita (LC) condition but are independent variables. The action can be written in terms of the two {\textit{distortion tensors}} 
\be \mathcal{C}_{ \mu\,\,\,\nu}^{\,\,\,\rho}\equiv\Gamma_{ \mu\,\,\,\nu}^{\,\,\,\rho}-\left.\Gamma_{ \mu\,\,\,\nu}^{\,\,\,\rho}(g)\right|_{LC},\,\,\,\,\,\,\widetilde{\mathcal{C}}_{ \mu\,\,\,\nu}^{\,\,\,\rho}\equiv\widetilde{\Gamma}_{ \mu\,\,\,\nu}^{\,\,\,\rho}-\left.\Gamma_{ \mu\,\,\,\nu}^{\,\,\,\rho}(f)\right|_{LC}\,,\ee
as
\begin{align}
\mathcal{S} =\int  {\rm d}^4x \Big\{& m_g^2\sqrt{-g}g^{ \mu\nu}\mathcal{R}_{ \mu\nu}(\mathcal{C})+m_f^2\sqrt{-f}f^{ \mu\nu}\mathcal{R}_{ \mu\nu}(\widetilde{\mathcal{C}}) \nonumber
\\& + 2m_g^2m^2\sqrt{-g}V\left(\sqrt{\Delta}\right)\Big\}\,.
\end{align}
The curvature scalars can be written in terms of the corresponding standard metric Ricci\footnote{The antisymmetrization of indices is defined as $T_{[\m\n]}=\frac{1}{2}\left(T_{\m\n}-T_{\n\m} \right)$.
} scalars as
\begin{align}
\label{CURV1}
\mathcal{R}(g,\mathcal{C})&=R(g)+2D_{[\mu}(g)C_{ \nu]}^{\,\,\,\mu\nu}+\mathcal{C}_{ \mu\,\,\,\,\lambda}^{\,\,\,\,\mu}\mathcal{C}_{ \nu}^{\,\,\,\lambda\nu}-\mathcal{C}_{ \nu\,\,\,\,\lambda}^{ \,\,\,\,\mu}\mathcal{C}_{ \mu}^{\,\,\,\lambda\nu}, 
\\[0.2cm]\mathcal{R}(f,\widetilde{\mathcal{C}})&=R(f)+2D_{[\mu}(f)\widetilde{\mathcal{C}}_{ \nu]}^{\,\,\,\mu\nu}+\widetilde{\mathcal{C}}_{ \mu\,\,\,\,\lambda}^{\,\,\,\,\mu}\widetilde{\mathcal{C}}_{ \nu}^{\,\,\,\lambda\nu}-\widetilde{\mathcal{C}}_{ \nu\,\,\,\,\lambda}^{ \,\,\,\,\mu}\widetilde{\mathcal{C}}_{ \mu}^{\,\,\,\lambda\nu},
\label{CURV2}
\end{align}
where the covariant derivatives are taken with respect to the corresponding Levi-Civita connections.
Varying the action with respect to $\mathcal{C}$ and $\widetilde{\mathcal{C}}$ we obtain that $\mathcal{C}_{ \mu\,\,\,\,\nu}^{\,\,\,\,\rho}=\delta_{ \mu}^{ \rho}Q_{ \nu}$ and $\widetilde{\mathcal{C}}_{ \mu\,\,\,\,\nu}^{\,\,\,\,\rho}=\delta_{ \mu}^{ \rho}\widetilde{Q}_{ \nu}$, where $Q_{ \nu}$ and $\widetilde{Q}_{ \nu}$ are arbitrary vectors. Substituting this solution into the curvature scalar expressions cancels out all extra terms and reduces them into the corresponding Ricci scalars. As a result the action returns into its familiar form~\eqref{ACT-0}. Therefore, the \textit{bimetric-affine}  formulation of the Einstein-Hilbert gravity is entirely equivalent to the standard bimetric formulation.

\vspace{-0.1cm}
\section{Quadratic Gravity}\label{Q_BAG}
\vspace{-0.1cm}
As we have explained in the introduction there is sufficient evidence that quantum corrections of matter fields coupled to gravity will generate modifications to the Einstein-Hilbert action. The simplest of these corrections corresponds to quadratic curvature terms and in particular to quadratic Ricci scalar terms, being safe from the point of view of not introducing any new ghostlike degrees of freedom. In our case of a bimetric theory such terms would be $R^2(g)$ or $R^2(f)$ or even a mixed term $R(g)R(f)$. In what follows we shall focus on the simplest case of just a $R^2(g)$ term. Thus, we consider the action
\begin{align}
\mathcal{S}= \int &{\rm d}^4x \,\Big\{ m_g^2\sqrt{-g}R(g)+m_f^2\sqrt{-f}R(f) \nonumber
\\& +\frac{m_g^2}{2\widetilde{m}^2}\sqrt{-g}R^2(g)\,+2m_g^2{m^2}\sqrt{-g}V\left(\sqrt{\Delta}\right)\Big\}\,.{\label{ACT-1}}
\end{align}
This can be rewritten in terms of an auxiliary scalar $\chi$ as
\begin{align}
\mathcal{S}= \int &{\rm d}^4x\,\Big\{m_g^2\sqrt{-g}\left(1+\frac{\chi}{\widetilde{m}^2}\right)R(g)+m_f^2\sqrt{-f}R(f)\nonumber
\\& -\frac{m_g^2}{2\widetilde{m}^2}\sqrt{-g}\chi^2+2m_g^2{m^2}\sqrt{-g}V\left(\sqrt{\Delta}\right)\Big\}\,.{\label{ACT-1b}}
\end{align}
The \textit{bimetric-affine} version of the action~\eqref{ACT-1} reads\footnote{The Riemann curvature tensor is defined as
$  \tns{\cal{R}}{_\m_\n^\r_\s}\,\equiv\,\partial_{ \mu}\tns{{\Gamma}}{_\n^\r_\s}-\partial_{ \nu}\tns{{\Gamma}}{_\m^\r_\s}+\tns{{\Gamma}}{_\m^\r_\l}\tns{{\Gamma}}{_\n^\l_\s}-\tns{{\Gamma}}{_\n^\r_\l}\tns{{\Gamma}}{_\m^\l_\s}$ while the Ricci scalar is given by $\cal{R}={\cal{R}}_{ \mu\nu}^{\,\,\,\,\,\mu\nu}$.}
\begin{align}
\mathcal{S}= &\int  {\rm d}^4x\,\Big\{m_g^2\sqrt{-g}{\cal{R}}(g,\mathcal{C})+m_f^2\sqrt{-f}{\cal{R}}(f,\widetilde{\mathcal{C}})\nonumber
\\& + \frac{m_g^2}{2\widetilde{m}^2}\sqrt{-g}\mathcal{R}^2(g,\mathcal{C})+2m_g^2{m^2}\sqrt{-g}V\left(\sqrt{\Delta}\right)\Big\}\,.
\label{ACT-2}
\end{align}
Note that this is not the most general \textit{bimetric-affine} quadratic action of scalar curvature invariants that could be written. Apart from the terms ${\cal{R}}^2(f,\widetilde{\mathcal{C}})$ and ${\cal{R}}(g,\mathcal{C}){\cal{R}}(f,\widetilde{\mathcal{C}})$ that were left aside, we could also have linear or quadratic terms of the parity-odd {\textit{Holst invariant}}\footnote{ The \textit{Holst invariant} is given by the contraction of the Levi-Civita antisymmetric symbol with the Riemann tensor, \textit{i.e.}.  $\widetilde{\cal{R}}=(-g)^{-1/2}\tns{\epsilon}{^\m^\n^\r^\s}\tns{\cal{R}}{_\m_\n_\r_\s}=2(-g)^{-1/2}\epsilon^{ \mu\nu\rho\sigma}\left(D_{ \mu}\mathcal{C}_{ \nu\rho\sigma}+\mathcal{C}_{ \mu\rho\lambda}\mathcal{C}_{ \nu\,\,\,\,\,\sigma}^{\,\,\,\lambda}\right)$.} $\widetilde{\cal{R}}$. These terms are known to introduce extra dynamical degrees of freedom. We also leave such terms aside and concentrate on the simplest of all quadratic cases of just an ${\cal{R}}^2(g,\mathcal{C})$ term. The equivalent auxiliary scalar form of the action~\eqref{ACT-2} is 
\begin{align}
\mathcal{S}= \int  &{\rm d}^4x\,\Big\{m_g^2\sqrt{-g}\left(1+\frac{\chi}{\widetilde{m}^2}\right){\cal{R}}(g,\mathcal{C})+m_f^2\sqrt{-f}{\cal{R}}(f,\widetilde{\mathcal{C}}) \nonumber
\\& -\frac{m_g^2}{2\widetilde{m}^2}\sqrt{-g}\chi^2+2m_g^2{m^2}\sqrt{-g}V\left(\sqrt{\Delta}\right)\Big\}\,.
\label{ACT-3}
\end{align}
Using the expressions~\eqref{CURV1} and~\eqref{CURV2} we write the action as
\begin{align}
&\mathcal{S}=\int{\rm d}^4x\,\Big\{\sqrt{-g}m_g^2\left(1+\frac{\chi}{\widetilde{m}^2}\right)R(g)+\sqrt{-f}m_f^2R(f) \nonumber
\\& +\sqrt{-g}m_g^2\left(1+\frac{\chi}{\widetilde{m}^2}\right)\left(2D_{[\mu}(g)\mathcal{C}_{ \nu]}^{\,\,\,\mu\nu}+\mathcal{C}_{ \mu\,\,\,\,\lambda}^{\,\,\,\,\mu}\mathcal{C}_{ \nu}^{\,\,\,\lambda\nu}-\mathcal{C}_{ \nu\,\,\,\,\lambda}^{ \,\,\,\,\mu}\mathcal{C}_{ \mu}^{\,\,\,\lambda\nu}\right)\nonumber
\\& +\sqrt{-f}m_f^2\left(2D_{[\mu}(f)\widetilde{\mathcal{C}}_{ \nu]}^{\,\,\,\mu\nu}+\widetilde{\mathcal{C}}_{ \mu\,\,\,\,\lambda}^{\,\,\,\,\mu}\widetilde{\mathcal{C}}_{ \nu}^{\,\,\,\lambda\nu}-\widetilde{\mathcal{C}}_{ \nu\,\,\,\,\lambda}^{ \,\,\,\,\mu}\widetilde{\mathcal{C}}_{ \mu}^{\,\,\,\lambda\nu}\right) \nonumber
\\&  -\frac{m_g^2}{2\widetilde{m}^2}\sqrt{-g}\chi^2+2m_g^2{m^2}\sqrt{-g}\,V\left(\sqrt{\Delta}\right)\,\Big\}\,.
\end{align}
Solving for $\widetilde{\mathcal{C}}_{ \mu\nu\rho}$ gives the same result as in the Einstein-Hilbert case and ultimately reduces ${\cal{R}}(f,\widetilde{\mathcal{C}})$ to just $R(f)$. Solving for $\mathcal{C}_{ \mu\nu\rho}$ on the other hand 
leads to an equation
\be
 \delta_{ \beta}^{ \alpha}\tns{\mathcal{C}}{_\n_\g^\n}+\delta_{ \gamma}^{ \alpha}\tns{\mathcal{C}}{_\n^\n_\b}-\tns{\mathcal{C}}{_\b_\g^\a}-\tns{\mathcal{C}}{_\g^\a_\b} =  2\partial_{ [\beta}\ln(1+\chi/\widetilde{m}^2)\delta_{ \gamma]}^{ \alpha}\,,
\ee
which has a solution
\be
\mathcal{C}_{ \mu\nu\rho}=g_{ \mu[\nu}\partial_{ \rho]}\ln(1+\chi/\widetilde{m}^2)\,.
\ee

Substituting back into the action we get it into the form
\begin{align}
& \mathcal{S}=\int{\rm d}^4x\,\Big\{\sqrt{-g}m_g^2\left(1+\frac{\chi}{\widetilde{m}^2}\right)R(g)+\sqrt{-f}m_f^2R(f) 
\\& +\frac{3}{4\widetilde{m}^2}\sqrt{-g}\frac{(\nabla\chi)^2}{\left(1+\frac{\chi}{\widetilde{m}^2}\right)} -\frac{m_g^2}{2\widetilde{m}^2}\sqrt{-g}\chi^2+2m_g^2{m^2}\sqrt{-g}V\left(\sqrt{\Delta}\right)\Big\}\,.\nonumber {\label{ACT-4}}
\end{align}
Next, we consider the Weyl rescaling of the $g_{ \mu\nu}$ metric
\be g_{ \mu\nu}\,\rightarrow\,\left(1+\frac{\chi}{\widetilde{m}^2}\right)^{-1}g_{ \mu\nu}\,,\ee 
which transforms the action to the Einstein frame as
\begin{align}
 \mathcal{S}=\int &{\rm d}^4x\,\Bigg\{m_g^2\sqrt{-g}R(g)+m_f^2\sqrt{-f} R(f) \nonumber
 \\&  -\frac{m_g^2}{2\widetilde{m}^2}\sqrt{-g}\frac{\chi^2}{\left(1+\frac{\chi}{\widetilde{m}^2}\right)^2} \nonumber
 \\& +\frac{2m_g^2{m^2}}{\left(1+\frac{\chi}{\widetilde{m}^2}\right)^2}\sqrt{-g}V\left((1+\chi/\widetilde{m}^2)^{1/2}\sqrt{\Delta}\right)\Bigg\}\,.
 {\label{ACT-5}}
\end{align}
Note that the kinetic term contribution generated by the Weyl rescaling exactly cancels the existing $\chi$-kinetic term, ending up with a nondynamical $\chi$ field in the Einstein frame.
Due to the property $e_n\left(\sqrt{\omega x}\right)=\omega^{n/2}e_n\left(\sqrt{x}\right)$, the detailed Weyl-rescaled potential term reads
\be 2m_g^2{m^2}\sqrt{-g}\sum_{n=0}^4\beta_ne_n\left(\sqrt{\Delta}\right) (1+\chi/\widetilde{m}^2)^{n/2-2}\,.\ee
Varying the action with respect to $\chi$ we obtain the equation
\be \frac{\chi}{\widetilde{m}^2}\,=\,\frac{2{m^2}}{\widetilde{m}^2}\sum_n\left(\frac{n}{2}-2\right)\beta_n\,e_n\left(\sqrt{\Delta}\right)\left(1+\frac{\chi}{\widetilde{m}^2}\right)^{n/2}\,.{\label{NONLINEARSOL}}\ee
The solution to this equation, substituted back into the action, gives the complete nonlinear bimetric theory in terms of the two metric tensors $g_{ \mu\nu}$ and $f_{ \mu\nu}$.
Nevertheless, as in the case of Einstein-Hilbert bimetric theory~\cite{Hassan:2011zd}, we must investigate whether there are suitable values of the parameters $\beta_n$ that allow for the correct ghost-free spectrum. Therefore, we should next consider the linearized limit of this theory.

\vspace{-0.1cm}
\section{Linearization of the \textit{Bimetric-Affine} Quadratic Action}\label{LIN}
\vspace{-0.1cm}

In order to identify the spectrum arising from the quadratic action
\eqref{ACT-5} and derive the necessary conditions for a ghost-free spectrum we must consider a linear approximation, keeping at most quadratic terms of the fields  in the same fashion as in the case of the Einstein-Hilbert bimetric action. 
 We first linearize the action with respect to the auxiliary $\chi$  keeping at most quadratic orders. The action~\eqref{ACT-5} takes the form
\begin{align}
 \mathcal{S}=\int {\rm d}^4x\,\Bigg\{& m_g^2\sqrt{-g}R(g)+m_f^2\sqrt{-f} R(f) -\frac{m_g^2}{2\widetilde{m}^2}\sqrt{-g}\chi^2 \nonumber
 \\& +2m_g^2{m^2}\sqrt{-g}\sum_{n=0}^4\beta_n\, \k_n\,e_n\left(\sqrt{\Delta}\right)\Bigg\}\,,
 {\label{ACT-6}}
\end{align}
with
\be
\k_n = 1 + \frac{\chi}{\widetilde{m}^2} \left(\frac{n}{2}-2 \right)+ \frac{\chi^2}{2\widetilde{m}^4}\left(\frac{n}{2}-2 \right)\left(\frac{n}{2}-3 \right)\,.
\ee
Variation of~\eqref{ACT-6} with respect to $\chi$ yields the solution\footnote{Note that~\eqref{LINEARSOL} is the same as the one resulting from the linearization of~\eqref{NONLINEARSOL}.}
\be \frac{\chi}{\widetilde{m}^2}=\frac{2\frac{{m^2}}{\widetilde{m}^2}\sum_n\beta_ne_n\left(\sqrt{\Delta}\right)(n/2-2)}{1-2\frac{{m^2}}{\widetilde{m}^2}\sum_n\beta_ne_n\left(\sqrt{\Delta}\right)(n/2-2)(n/2-3)}\,.{\label{LINEARSOL}}\ee

Varying with respect to $g_{\m\n}$ and $f_{\m\n}$ we obtain the following Einstein equations
\begin{subequations}
\begin{align}
\label{eq:eomg}
R_{\m\n}(g)-\frac{1}{2}g_{\m\n}R(g)+\frac{\chi^2}{4\widetilde{m}^2} g_{\m\n} - {m^2} \sum_{n=0}^3 \b_n \k_n V^{(n)}_{\m\n} &= 0\,, 
\\R_{\m\n}(f)-\frac{1}{2}f_{\m\n}R(f) - \frac{{m^2}m_g^2}{{m_f^2}} \sum_{n=1}^4 \b_n \k_n \widetilde{V}^{(n)}_{\m\n} &= 0\,.
\label{eq:eomf}
\end{align}
\end{subequations}
The analytic forms of the matrices $V^{(n)}_{\m\n}$ and $\widetilde{V}^{(n)}_{\m\n}$ are given in~\ref{appendix_e_n}.

Next we proceed to complete the linearization by approximating the dynamical fields as
\be g_{ \mu\nu}\,\approx\,\bar{g}_{ \mu\nu}+h_{ \mu\nu},\,\,\,\,\,\,\,f_{ \mu\nu}\,\approx\,\bar{g}_{ \mu\nu}\,+\,l_{\mu\nu}\,.\ee
An important class of solutions  are the proportional solutions $\bar{f}_{\m\n} = c^2 \bar{g}_{\mu\n}$~\cite{Hassan:2012wr} in which we can set $c^2=1$ without loss of generality. Substituting this ansatz the equations of motion are reduced to
\begin{subequations}
\begin{align}
\label{eq:eoms_prop_0}
R_{\m\n}(\bar{g})-\frac{1}{2}\bar{g}_{\m\n}R(\bar{g})
+\Lambda_g \bar{g}_{\m\n} &= 0\,,
\\R_{\m\n}(\bar{g})-\frac{1}{2}\bar{g}_{\m\n}R(\bar{g}) +\Lambda_f \bar{g}_{\m\n}&= 0\,,
\label{eq:eoms_prop}
\end{align}
\end{subequations}
with
\begin{subequations}
\begin{align}
\Lambda_g &= \frac{\chi^2}{4 \widetilde{m}^2} -{m^2} \left( \b_0 \k_0 +3\b_1 \k_1 +3\b_2 \k_2 +\b_3 \k_3 \right)\,,
\\ \Lambda_f &=  -\frac{{m^2} m_g^2}{m_f^2}\left( \b_1 \k_1 +3\b_2 \k_2 +3\b_3 \k_3+\b_4 \k_4 \right)\,.
\end{align}
\end{subequations}
Note that the auxiliary $\chi$ is substituted using equation~\eqref{LINEARSOL} in which the polynomials $e_n$ (for the proportional solutions) are in turn $1,4,6,4,1$. Its corresponding expression reads
\be \frac{\chi}{\widetilde{m}^2}\,=\,\frac{\frac{4m^2}{\widetilde{m}^2}\left(\beta_0+3\beta_1+3\beta_2+\beta_3\right)}{\frac{6m^2}{\widetilde{m}^2}\left(2\beta_0+5\beta_1+4\beta_2+\beta_3\right)-1}\,.\ee
Consistency between the equations of motion~\eqref{eq:eoms_prop_0}-\eqref{eq:eoms_prop} requires $\Lambda_g=\Lambda_f$, which fixes one of the $\b_n$ parameters.

After quite a bit of algebra we arrive at the linearized action, which has the same general form as in the Einstein-Hilbert bimetric case given by equation~\eqref{eq:lin_act}, namely \newpage
\begin{align}
\label{eq:act_fin}
\mathcal{S} = \int {\rm d}^4x &\sqrt{-\bar{g}} \bigg\{\frac{m_g^2}{4}h_{ \mu\nu}\tns{\mathcal{\mathcal{E}}}{^\m^\n_\r_\s}h^{ \rho\sigma}\,+\,\frac{m_f^2}{4}l_{ \mu\nu}\tns{\mathcal{\mathcal{E}}}{^\m^\n_\r_\s}l^{ \rho\sigma} \nonumber
\\&  +\widetilde{B} \left[ \left( \tns{h}{_\m_\n}-\tns{l}{_\m_\n}\right)^2 + \widetilde{C}\left( h-l\right)^2 \right] \nonumber 
\\& +  \widetilde{A}_g \left( h_{\m\n} h^{\m\n} -\frac{1}{2}h^2 -2h -4\right) \nonumber
\\&+  \widetilde{A}_f \left( l_{\m\n} l^{\m\n} -\frac{1}{2}l^2 -2l-4 \right)  \bigg\}\,.
\end{align}
The appearing parametric coefficients $\widetilde{A}_g,\,\widetilde{A}_f,\,\widetilde{C}$ and $\widetilde{B}$ are given by complicated expressions of the parameters $\beta_n$, $m_g\,,m_f\,$ and $\l={m^2}/\widetilde{m}^2$ in~\ref{appendix}. Note also that the parameters of the third and forth line of~\eqref{eq:act_fin} are related to the corresponding cosmological constants by the relations $\Lambda_g = 2\widetilde{A}_g/m_g^2$ and $\Lambda_f = 2\widetilde{A}_f/m_f^2$.  Again the spectrum can consist of a massless spin-$2$ and a massive spin-$2$ particle, without the presence of any ghosts, provided the following conditions on the parameters are met, namely
\be \Lambda_g=\Lambda_f\,=\,\Lambda,\,\,\,\,\,\widetilde{C}=-1\quad \text{and} \quad \widetilde{B}<0\,,{\label{constraints}}\ee
with or without the additional requirement of a vanishing cosmological constant $\Lambda=0$.
The equation $\widetilde{C}=-1$ is most easily {\textit{``solved"}} with respect to the parameter $\l$ as
\be
\l_{\star}=\frac{\b_1+2\b_2+\b_3}{2\b_0(\b_1+4\b_2+3\b_3)+6\b_1\b_2+8\b_1\b_3+2\b_2\b_3}\,.
\label{eq:newlambda}
\ee
This specific constraint on the parameters is adequate to remove the ghost degree of freedom from the linearized theory.
The ``trivial" case with $\l=0$ discussed in section~\ref{BAG} is also a solution. Substituting, $\l_{\star}$ into~\eqref{eq:A1}-\eqref{eq:B1_app} we obtain the simplified expressions
\begin{subequations}
\begin{align}
\widetilde{A}_g\left|_{\l=\l_{\star}} \right. =& -m_g^2 {m^2} \big[21\b_1^2 +36\b_2^2+5\b_3^2+26\b_2\b_3 \nonumber
\\& +18\b_1(3\b_2+\b_3) +2\b_0(3\b_1+4\b_2+\b3)\big]\nonumber
\\& \times \big[4(5\b_1+8\b_2+3\b_3) \big]^{-1}\,,
\label{eq:A1_lgf}
\\[0.1cm]  \nonumber
\widetilde{A}_f\left|_{\l=\l_{\star}} \right. =& -m_g^2 {m^2} \big[7\b_1^2 + 36\b_2^2 +15\b_3^2 +34\b_1\b_2  \nonumber
\\& +54\b_2\b_3 +2\b_4(8\b_2+3\b_3) +10\b_1(3\b_3+\b_4) \big]\nonumber
\\& \times \big[4(5\b_1+8\b_2+3\b_3) \big]^{-1}\,,
\label{eq:A2_lgf}
\\[-0.4cm]  \nonumber
\\ \widetilde{B}\left|_{\l=\l_{\star}} \right.=&  m_g^2 {m^2}(\b_1+2\b_2+\b_3)^2
\\& \times(35\b_1+48\b_2+15\b_3)\big[8(5\b_1+8\b_2+3\b_3)^2\big]^{-1}\,.
\label{eqB1_lgf}\nonumber
\end{align}
\end{subequations}
Note here that the $\widetilde{B}\left|_{\l=\l_{\star}} \right.$ no longer depends on the parameter $\b_0$. The Fierz-Pauli mass reads 
\begin{align} 
m_{\rm FP}^2 =& -\frac{{m^2}(m_g^2+m_f^2)}{{m_f^2}}(\b_1+2\b_2+\b_3)^2
\\[-0.2cm]& \times(35\b_1+48\b_2+15\b_3)\big[2(5\b_1+8\b_2+3\b_3)^2\big]^{-1}\,.
\label{eq:FP_lgf}\nonumber
\end{align}

\section{Parameter Space}{\label{Parameter}}
Assuming the values of parameters for the minimal massive model introduced in~\cite{Hassan:2011vm}, \textit{i.e.} $\b_0=3\,,\, \b_1=-1\,, \,\b_2=0\,,\,\b_3=0\,,\, \b_4=1$, the functions $\widetilde{A}_g, \widetilde{A}_f$ vanish, while $\widetilde{C}=-\frac{2-3\l}{2-12\l}$. It is evident that for this choice of the parameters the quadratic theory will necessarily be plagued with the ghost degree of freedom, since $\widetilde{C}=-1$ only if $\l=0$, \textit{i.e.} in the absence of the  $\mathcal{R}^2$ term.

The parameter space of the bimetric theory\footnote{Cosmological constraints on the parameters of bimetric (and massive) gravity have been analyzed extensively in the literature~\cite{DAmico:2011eto, vonStrauss:2011mq, Koennig:2013fdo, Volkov:2011an, Volkov:2012zb, Akrami:2012vf, Solomon:2014dua, Koennig:2014ods, Mortsell:2018mfj, Lindner:2020eez, DeFelice:2014nja, Akrami:2015qga, Luben:2018ekw, Luben:2020xll, Hogas:2021lns, Caravano:2021aum, Hogas:2021fmr, Hogas:2021saw, Bassi:2023ymf}.} is spanned by our set of physical parameters  $m_g,\,\alpha=m_f/m_g$, and $\lambda$ as well as the model parameters $\beta_n$. The enforcement of the constraints~\eqref{constraints} results in fixing two of the $\beta$'s. In what follows we shall analyze the various cases of the resulting viable bimetric models.

\subsection{\textbf{The $\mathbf{\beta_2=\beta_3=0 }$ case.}}

We proceed analyzing first the case of models corresponding to the choice of $\beta_2=\beta_3=0$. In this case the $\widetilde{C}=-1$ constraint fixes $\beta_0$ to be $\beta_0=\frac{1}{2\lambda}$. 
The corresponding Fierz-Pauli mass expression is 
\be m_{\rm FP}^2\,=\,-\frac{7}{10}\beta_1\left(\frac{m}{m_f}\right)^2M_P^2\,,\ee 
which requires a negative value for the free parameter $\beta_1$. The parameter $\beta_4$ is also fixed by the $\Lambda_g=\Lambda_f$ condition, although it does not play any role, since it does not appear in the expression for $m_{\rm FP}$. The common value of the cosmological constant is
\be \Lambda\,=\,-\frac{21m^2}{10}\left(\beta_1+\frac{1}{7\lambda}\right)\,.\ee
The so-called {\textit{Higuchi bound}}\footnote{If this bound is violated, the helicity-0 mode of the massive spin-2 field develops the wrong sign in its kinetic term.}~\cite{Higuchi:1986py, Higuchi:1989gz} $m_{\rm FP}^2>2\Lambda/3$, after replacing $M_P^2=m_g^2(1+\alpha^2)$, leads to the following restriction on $\beta_1$:
\be 
\left\{ 
    \begin{array}{lr}
      \b_1 > \frac{2\a^2}{7\l(1-\a^2)}, \quad  \text{if}\,\,\, \a>1\\
       \b_1< \frac{2\a^2}{7\l(1-\a^2)}, \quad \text{if}\,\,\, \a<1\,.
    \end{array}
\right.
\ee
Of course, the $\alpha<1$ case is always met for $\beta_1<0$.

Enforcing the constraint of vanishing cosmological constant fixes $\beta_1$ to be $\beta_1=-\frac{1}{7\lambda}$ and gives the following expression for the Fierz-Pauli mass
\be m_{\rm FP}^2\,=\,\frac{1}{10\lambda}\left(\frac{m}{m_f}\right)^2M_P^2\,.\ee
Assuming that $m=m_f$, we may rewrite $m_{\rm FP}$ in terms of the mass scale of the quadratic correction $\widetilde{m}$ as
\be m_{\rm FP}^2=\frac{M_P^2}{10\lambda}\,=\,\frac{\widetilde{m}^2}{10}(1+\alpha^{-2})\,.\ee

\subsection{\textbf{The $\mathbf{\beta_2, \beta_3 }$ nonzero case.}}
\begin{figure}[t!]  
\begin{center}
\includegraphics[width=0.495\textwidth]{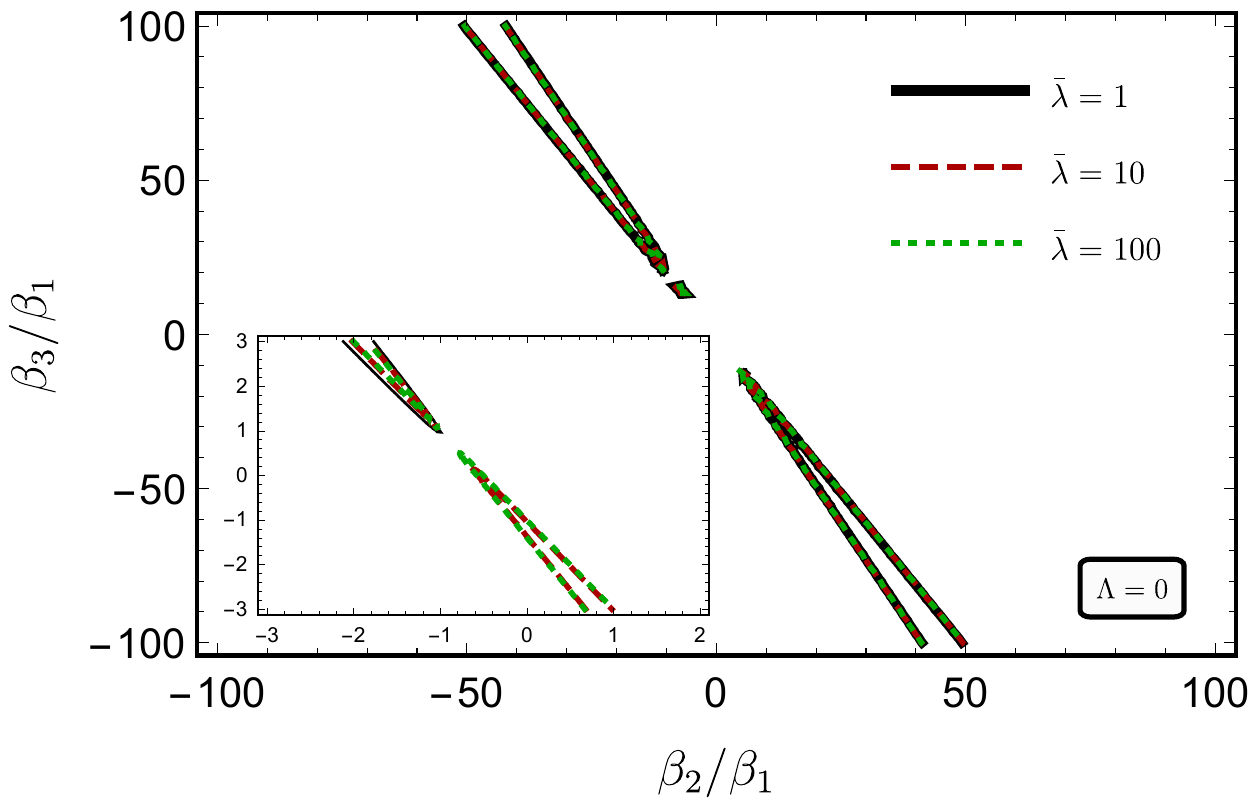}
\caption{The lines along which the equation~\eqref{eq:solLz2nd} is valid for various values of $\Bar{\l}$. Along these lines the cosmological constant vanishes. The insertion at the bottom left shows an enlargement of the plot for small values.}
\label{fig:2}
\end{center}
\end{figure}
\begin{figure*}[h]  
\begin{center}
\includegraphics[width=0.495\textwidth]{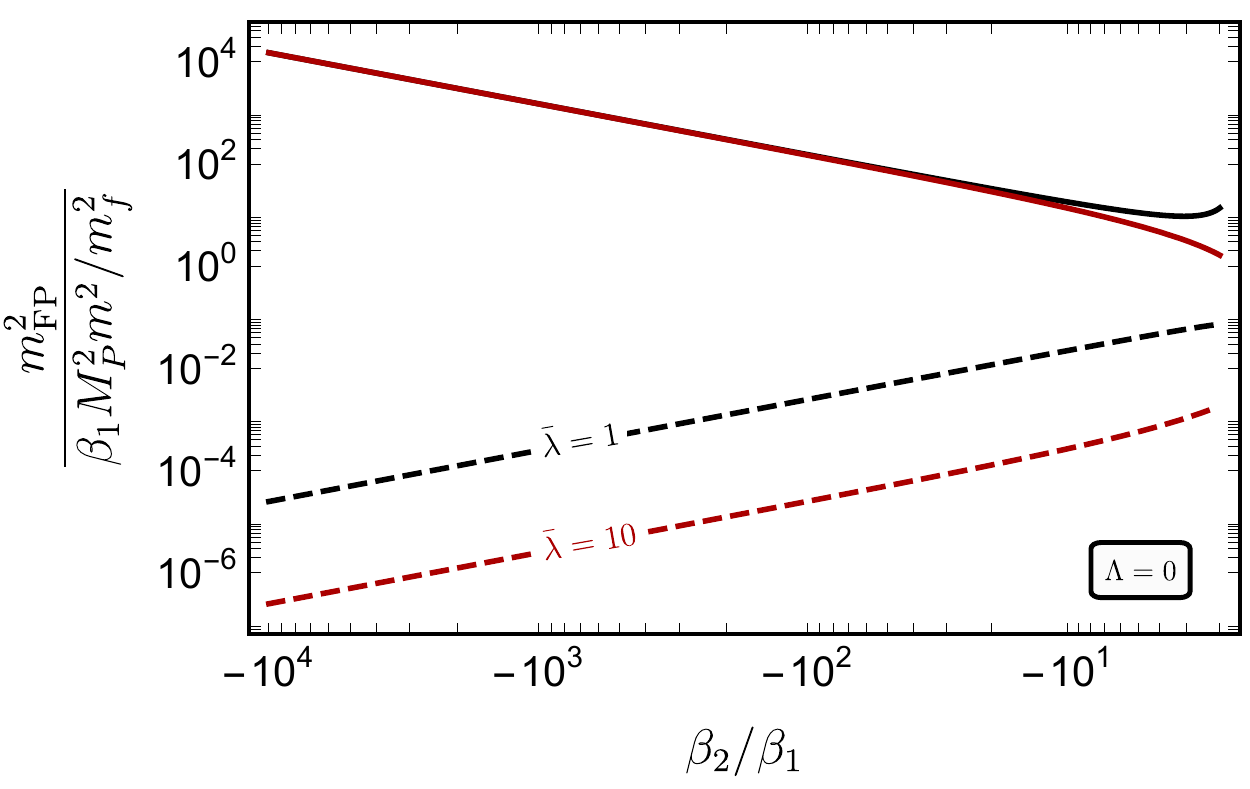}
\includegraphics[width=0.495\textwidth]{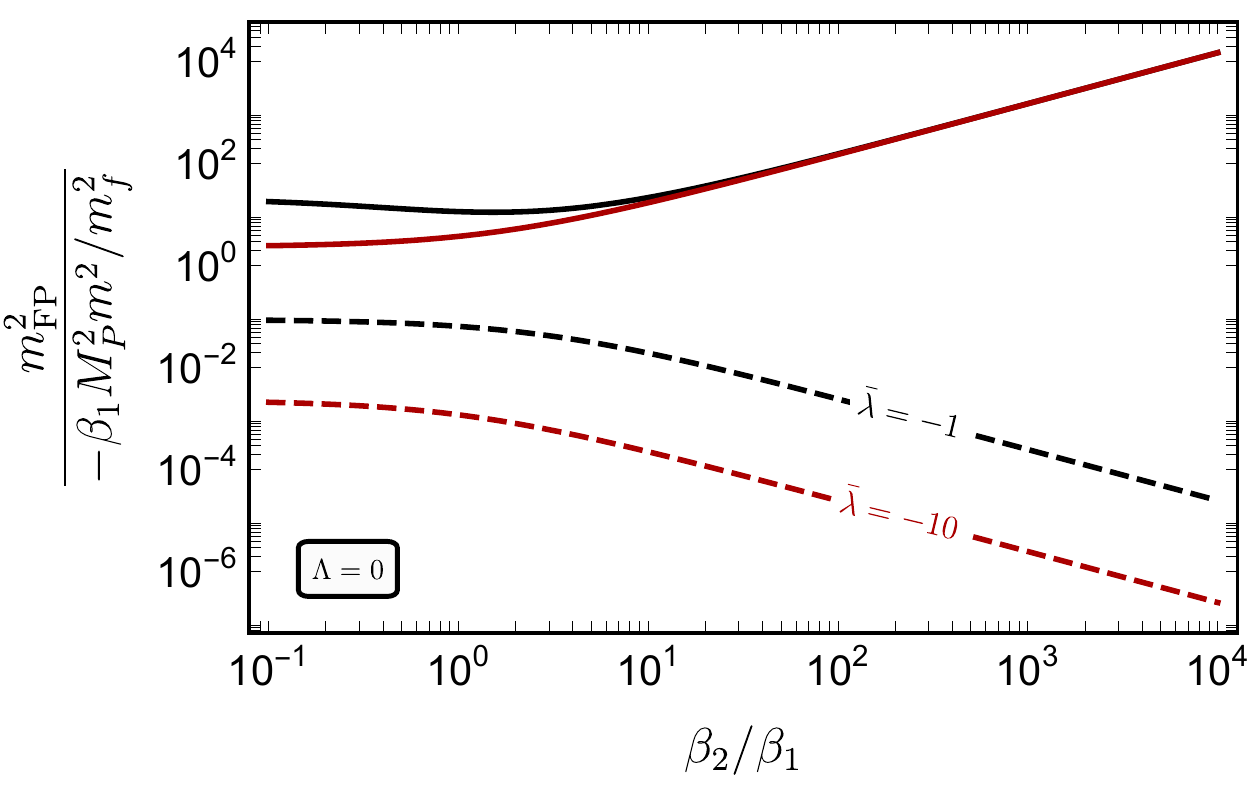}
\caption{ \textbf{Left:} The Fierz-Pauli mass in units of $\b_1 M_P^2m^2/m_f^2$ as a function of the ratio $\b_2/\b_1$ for the choice~\eqref{eq:b1signpos}. \textbf{Right:} The Fierz-Pauli mass in units of $-\b_1 M_P^2m^2/m_f^2$ as a function of the ratio $\b_2/\b_1$ for the choice~\eqref{eq:b1signneg}. In both panels the solid lines correspond to the plus sign solution of~\eqref{eq:solLz2nd}, while the dashed ones correspond to the minus sign. The same colors indicate the same $\Bar{\l}$. }
\label{fig:mfp}
\end{center}
\end{figure*}

For general nonzero $\beta_2,\,\beta_3$ we can invert \eqref{eq:newlambda} as
\be 
\beta_0 = \frac{\b_1+2\b_2+\b_3-2\l(3\b_1\b_2+4\b_1\b_3+\b_2\b_3)}{2\l(\b_1+4\b_2+3\b_3)}\,.
\label{BETA-0}
\ee 
Note that in the case $\beta_2=\beta_3=0$ we recover $\beta_0=\frac{1}{2\lambda}$. 
Next, using the constraint equation $\Lambda_g=\Lambda_f$ we also fix the parameter $\b_4$ as
\begin{align}
\beta_4&=\big[7\beta_1^2(3\alpha^2-1)+36\beta_2^2(\alpha^2-1)+5\beta_3^2(\alpha^2-3) \nonumber
\\& +2\beta_2\beta_3(13\alpha^2-27)+2\beta_1\beta_2(27\alpha^2-17)+6\beta_1\beta_3(3\alpha^2-5) \nonumber
\\& +2\beta_0(3\beta_1+4\beta_2+\beta_3)\alpha^2\big]\left[2({8}\beta_2+3\beta_3+5\beta_1)\right]^{-1}\,.
\end{align}
Actually $\beta_4$ does not play much of a role, since it does not appear in $m_{\rm FP}$. Considering the parameters $\beta_0$ and $\beta_4$ as fixed, the resulting models are parametrized in terms of $\beta_2,\,\beta_3$, and $\beta_1$. 

The resulting expressions for the Fierz-Pauli mass and the cosmological constant, written in terms of the rescaled parameters 
\be \bar{\lambda}\,=\,\lambda\beta_1,\,\,\,\bar{\beta}_{2,3}\,=\,\frac{\beta_{2,3}}{\beta_1}\,,\ee 
are
\begin{align}
\label{eq:mfp22nd}
m_{\rm FP}^2 =& -\b_1 M_P^2 \left(\frac{m}{m_f} \right)^2 (35+48\Bar{\beta}_2+15\Bar{\beta}_3)(1+2\Bar{\beta}_2+\Bar{\beta}_3)^2\nonumber
\\& \times \left[2(5+8\Bar{\beta}_2+3\Bar{\beta}_3)^2 \right]^{-1}\,,
\end{align}
and
\newpage
\begin{align}
\label{eq:Lambda2nd}
\Lambda =& -m^2 \b_1 (1+2\Bar{\beta}_2+\Bar{\beta}_3) \big( 3+4\Bar{\beta}_2+\Bar{\beta}_3+ \Bar{\l} (21+ 78\Bar{\beta}_2\nonumber
\\& +36\Bar{\beta}_3 + 72\Bar{\beta}_2^2+15\Bar{\beta}_3^2 +66\Bar{\beta}_2\Bar{\beta}_3 \big) \nonumber
\\& \times\Big[2\Bar{\l}(1+4\Bar{\beta}_2+3\Bar{\beta}_3)(5+8\Bar{\beta}_2+3\Bar{\beta}_3) \Big]^{-1}\,.
\end{align}
 In contrast to the $\b_2=\b_3=0$ case, in which the positivity of the Fierz-Pauli mass requires $\b_1<0$, in this case $m_{\rm FP}^2 >0$ is maintained for
\begin{subequations}
\be
\b_1 > 0 \quad \text{and} \quad 35+48\Bar{\beta}_2+15\Bar{\beta}_3 <0
\label{eq:b1signpos}
\ee
or
\be
\b_1 < 0 \quad \text{and} \quad 35+48\Bar{\beta}_2+15\Bar{\beta}_3 >0\,.
\label{eq:b1signneg}
\ee
\end{subequations}
Enforcing the constraint of vanishing cosmological constant reduces the number of free parameters to two. Thus, for $\Lambda=0$ we get the additional constraint in the form of a quadratic equation for $\bar{\beta}_3$ in terms of $\bar{\beta}_2$, namely
\begin{align} \label{eq:solLz2nd} & \Bar{\l} \left(21+ 78\Bar{\beta}_2+36\Bar{\beta}_3 + 72\Bar{\beta}_2^2+15\Bar{\beta}_3^2 +66\Bar{\beta}_2\Bar{\beta}_3\right) \nonumber
\\ & +3+4\Bar{\beta}_2+\Bar{\beta}_3=0\,.\end{align}
In figure~\ref{fig:2} we have plotted the solutions of this equation for various values of the parameter $\Bar{\l}$. As can be seen the solutions are not very sensitive on the choice of $\Bar{\l}$. Pairs of $(\bar{\beta}_2,\,\bar{\beta}_3)$ on this plot define different models with vanishing cosmological constant and a Fierz-Pauli mass given by \eqref{eq:mfp22nd}. Substituting the solutions of~\eqref{eq:solLz2nd},
\begin{align}
\label{eq:b3sol12}
\Bar{\beta}_3 =\frac{-1-6\Bar{\l}(6+11\Bar{\beta}_2)\pm \sqrt{1-108\Bar{\l}(1+\Bar{\beta}_2)+36\Bar{\l}^2(1+\Bar{\beta}_2)^2}}{30\Bar{\l}}\,,
\end{align}
into~\eqref{eq:mfp22nd} we obtain that
\be
\Bar{\beta}_2 \neq -\frac{1+41\Bar{\l}\pm \sqrt{\Bar{\l}^2+22\Bar{\l}+1}}{36\Bar{\l}}\,,
\ee
in order to avoid a vanishing Fierz-Pauli mass. The values $\Bar{\beta}_2 = -1 - \{0,1,2\}/4\Bar{\l}$ are also excluded since for these values the denominator of~\eqref{eq:Lambda2nd} vanishes.  For $\b_2=0$ the solution of~\eqref{eq:solLz2nd} with the minus sign gives $\b_3=0$ for $\Bar{\l}=-1/7$ as it should be. The other solution does not coincide with the $\b_2=\b_3=0$ case for any real value of $\Bar{\l}$. Figure~\ref{fig:mfp}  illustrates the Fierz-Pauli mass in units of the overall factor
$\pm\b_1 M_P^2m^2/m_f^2$. The left panel coincides with the choice~\eqref{eq:b1signpos}. This choice enforces $\Bar{\l}=\l\b_1$ to be positive, since the parameter $\l=m^2/\widetilde{m}^2$ is positive by definition. Also, for this choice we have that $\b_2<0$, while $\b_3>0$. In the right panel of the same figure we plot the choice~\eqref{eq:b1signneg}. This choice in turn, enforces $\Bar{\l}<0$, while again $\b_2<0$ and $\b_3>0$. In both panels the solid lines correspond to the  $``+"$-sign solution of~\eqref{eq:solLz2nd}, while the dashed ones to the $``-"$-sign one.

Having in mind the phenomenological motivations for considering a massive spin-$2$ particle, we should also see whether the parameter space allows its mass to be low enough so that it is phenomenologically interesting. Reading off from figure~\ref{fig:mfp} we see that for $|\bar{\lambda}|\sim \mathcal{O}(10)$ and $|\beta_2/\beta_1|\sim\mathcal{O}(1)$, we have $m_{\rm FP}^2/M_P^2\sim 10^{-3}|\beta_1|$, after choosing the redundant parameter $m=m_f$ . Thus, it seems that a relatively light $m_{\rm FP}$ could be maintained only for very small model values of $\beta_1$. For instance, taking $|\beta_1|\sim\mathcal{O}(10^{-25})$, we obtain $ m_{\rm FP}\sim\mathcal{O}(10\,{\rm TeV})$,  while $\widetilde{m}\sim \mathcal{O}(10^{5}\,{\rm GeV})$, if $m_f\simeq m_g \simeq M_P$. Of course, $m_{\rm FP}$ can be smaller at the price of even smaller values of the relevant potential parameters. Thus, although the possibility of a dark matter interpretation of the massive spin-2 is not excluded, this has to go along with rather small values of the interaction parameters of the theory.

\vspace{-0.1cm}
\section{Conclusions}\label{Conclusions}
\vspace{-0.1cm}
In the present article we considered the ghost-free bimetric theory of gravity~\cite{Hassan:2011zd} in a general \textit{bimetric-affine} framework where the connections associated to both Ricci scalars are independent variables. Furthermore, we considered quadratic curvature corrections to the standard Einstein-Hilbert type of action, expected to arise from matter fields quantum interactions. We focused on the simplest case of a quadratic Ricci scalar term associated to one of the metrics. The resulting theory, its spectrum consisting only on the standard massless graviton and a massive spin-$2$ field, was analyzed in its linearized limit and the constraints for the absence of a {\textit{Boulware-Deser}} ghost~\cite{Boulware:1972yco} were derived. We analyzed the constraints and solved the corresponding parameter equations, thus, identifying the parameter space that defines viable models. Our results show that you can extend bimetric gravity in the metric-affine framework including quadratic curvature terms, while maintaining the absence of ghosts, for a wide range of model parameters.

\vspace{-0.1cm}
\section*{Acknowledgments} 
\vspace{-0.1cm} 

\noindent   The work of IDG was supported by the Estonian Research Council grant SJD18. IDG wants to thank H.~Veerm\"ae, M.~Raidal and L.~Marzola for discussions. KT would like to thank C.~Bachas for illuminating discussions.  We thank also A.~Delhom  for communications.

\appendix
\section{The $V^{(n)}_{\m\n}$ and $\widetilde{V}^{(n)}_{\m\n}$ functions}\label{appendix_e_n}

The matrices $V_{ \mu\nu}^{(n)}$ appearing in the equations of motion~\eqref{eq:eomg},~\eqref{eq:eomf} are given by
\begin{align}
&V^{(0)}_{\m\n} = g_{\m\n}\,, \nonumber
\\ &V^{(1)}_{\m\n} = g_{\m\n} {\rm Tr}\left[X\right] - g_{\n\r} \tns{X}{^\r_\m}\,, \nonumber
\\ &V^{(2)}_{\m\n} = g_{\n\r} \left( \tns{{X^2}}{^\r_\m} -  {\rm Tr}\left[X\right]  \tns{X}{^\r_\m} \right) +\frac{g_{\m\n}}{2} \left( {\rm Tr}\left[X\right]^2 -{\rm Tr}\left[X^2\right]\right)\,, \nonumber
\\ &V^{(3)}_{\m\n} = -g_{\n\r}\left( \tns{{X^3}}{^\r_\m} - {\rm Tr}\left[X\right] \tns{{X^2}}{^\r_\m} +\frac{1}{2} \left(  {\rm Tr}\left[X\right]^2 -{\rm Tr}\left[X^2\right]\right)   \tns{X}{^\r_\m} \right)\nonumber
\\& \,\,\,\,\,\,\,\,+ \frac{g_{\m\n}}{6} \left( {\rm Tr}\left[X\right]^3 -3 {\rm Tr}\left[X\right] {\rm Tr}\left[X^2\right] + 2{\rm Tr}\left[X^3\right] \right)\,,
\end{align}
and
\begin{align}
\widetilde{V}^{(1)}_{\m\n} &= f_{\n\r} \tns{X}{^\r_\m}\,, \nonumber
\\ \widetilde{V}^{(2)}_{\m\n} &= f_{\n\r} \left({\rm Tr}\left[X\right]  \tns{X}{^\r_\m} - \tns{{X^2}}{^\r_\m}\right) \,, \nonumber
\\ \widetilde{V}^{(3)}_{\m\n} &= f_{\n\r}\left( \tns{{X^3}}{^\r_\m}  - {\rm Tr}\left[X\right]  \tns{{X^2}}{^\r_\m}  +\frac{1}{2} \left( {\rm Tr}\left[X\right]^2 - {\rm Tr}\left[X^2\right]\right)  \tns{X}{^\r_\m}\right)\,, \nonumber
\\ \widetilde{V}^{(4)}_{\m\n} &= f_{\m\n}  \,,
\end{align}
where $X=\sqrt{\Delta}$, defined in ({\ref{DELTA}}). 

\section{Formulas}\label{appendix}
The functions involved in~\eqref{eq:lin_act} are given by
\begin{align} 
\widetilde{A}_g =& -m_g^2 {m^2} (\b_0+3(\b_1+\b_2)  +\b_3)\big[ 6\l^2 (\b_0+3(\b_1+\b_2)  \nonumber
\\& + \b_3)(16\b_0+35\b_1+24\b_2+5\b_3) - 2\l (10\b_0  \nonumber
\\&+27\b_1+24\b_2 +7\b_3) +1\big] \big[2(1-6\l (2\b_0+5\b_1 \nonumber
\\& +4\b_2 +\b_3))^2 \big]^{-1}\,,
\label{eq:A1}
\\& \nonumber
\\ \widetilde{A}_f =& -m_g^2{m^2}\big[6\l^2\big(\b_0^2(17\b_1+56\b_2+63\b_3)+84\b_0\b_1^2 \nonumber
\\& + 6\b_4(2\b_0+5\b_1+4\b_2+\b_3)^2 +105\b_1^3 +216\b_2^3 + 15\b_3^3 \nonumber
\\& +342\b_0\b_1\b_2 +328\b_0\b_1\b_3 +216\b_0\b_2^2 +298\b_0\b_2\b_3 \nonumber
\\& + 60\b_0\b_3^2 +510\b_1^2\b_2 +429\b_1^2\b_3 +609\b_1\b_2^2 \nonumber
\\& +780\b_1\b_2\b_3+155\b_1\b_3^2 +351\b_2^2\b_3 +134\b_2\b_3^2\big) \nonumber
\\& -6\l \big(\b_0 (3\b_1+10\b_2+11\b_3+4\b_4) +7\b_1^2+18\b_2^2 \nonumber
\\& +5\b_3^2 +2\b_4(5\b_1+4\b_2+\b_3) +29\b_1\b_2 +28\b_1\b_3 \nonumber
\\& +25\b_2\b_3 \big)  +\b_1 +3(\b_2+\b_3)+\b_4\big]\big[2(1-6\l (2\b_0 \nonumber
\\& +5\b_1+4\b_2 +\b_3))^2 \big]^{-1}\,,
\label{eq:A2}
\\& \nonumber
\\\widetilde{B} =& m_g^2 {m^2} \big[2\l^2 \big(\b_0^2 (51\b_1 +112\b_2 +63\b_3) +\b_0 (252\b_1^2 \nonumber
\\& +750\b_1\b_2 +360\b_1\b_3 +432\b_2^2 +350\b_2\b_3+60\b_3^2)\nonumber
\\&+351\b_1^3 +27\b_1^2 (44\b_2+19\b_3)+3\b_1(429\b_2^2+55\b_3^2 \nonumber
\\& +328\b_2\b_3) +432\b_2^3 +15\b_3^3 +459\b_2^2\b_3+148\b_2\b_3^2\big) \nonumber
\\& -2\l (\b_0(9\b_1+20\b_2+11\b_3) +\b_1(63\b_2+30\b_3) \nonumber
\\&  +21\b_1^2 +(4\b_2+\b_3)(9\b_2+5\b_3))+\b_1+2\b_2+\b_3\big] \nonumber
\\& \times \big[4(1-6\l(2\b_0+5\b_1+4\b_2+\b_3))^2\big]^{-1} \label{eq:B1_app}\,,
\\& \nonumber
\\\widetilde{C} =&  \big[4\l^2(\b_0^2(159\b_1 +352\b_2+195\b_3)+\b_0(783\b_1^2 +1416\b_2^2 \nonumber
\\&  + 1172\b_2\b_3 +213\b_3^2 +12\b_1(198\b_2+95\b_3) )+945\b_1^3 \nonumber
\\& +9\b_1^2(401\b_2+175\b_3)+9\b_1(473\b_2^2 +340\b_2\b_3   \nonumber
\\& +59\b_3^2)+1296\b_2^3 +1389\b_2^2\b_3 +451\b_2\b_3^2+45\b_3^3) \nonumber
\\& -12\l^3(4\b_0^3(51\b_1+112\b_2+63\b_3)+\b_0^2(1521\b_1^2 \nonumber
\\& +4552\b_1\b_2 +2190\b_1\b_3 +2672\b_2^2+2200\b_2\b_3  \nonumber
\\ & +393\b_3^2) +2\b_0 (1890\b_1^3 +\b_1^2 (7143\b_2 +3090\b_3) \nonumber
\\& +2\b_1(3885\b_2^2 +2993\b_2\b_3+513\b_3^2)+2592\b_2^3  \nonumber
\\& +2762\b_2^2\b_3+895\b_2\b_3^2 +90\b_3^3)+3150\b_1^4
\nonumber
\\& +2880\b_1^3(5\b_2+2\b_3) +3\b_1^2(7467\b_2^2 +908\b_3^2 \nonumber
\\& +5464\b_2\b_3)+2\b_1(7308\b_2^3 +240\b_3^3 +7527\b_2^2\b_3 \nonumber
\\& +2396\b_2\b_3^2 )+3456\b_2^4 +30\b_3^4 +4536\b_2^3\b_3 \nonumber
\\& +2105\b_2^2\b_3^2 +416\b_2\b_3^3 )-\l(4\b_0(15\b_1 +32\b_2 \nonumber
\\& +17\b_3)+153\b_1^2 +6\b_1(76\b_2+35\b_3) \nonumber
\\& +276\b_2^2+224\b_2\b_3+41\b_3^2)+2(\b_1+2\b_2+\b_3)\big] \nonumber
\\& \times \big[2(6\l(2\b_0+5\b_1+4\b_2+\b_3)-1)(2\l^2(\b_0^2(51\b_1  \nonumber
\\& +112\b_2 +63\b_3)+\b_0(252\b_1^2+750\b_1\b_2+360\b_1\b_3  \nonumber
\\& +432\b_2^2 +60\b_3^2 +350\b_2\b_3)+351\b_1^3+27\b_1^2(44\b_2 \nonumber
\\& +19\b_3)+3\b_1(429\b_2^2 +328\b_2\b_3+55\b_3^2)+432\b_2^3 \nonumber
\\& +459\b_2^2\b_3+148\b_2\b_3^2+15\b_3^3) -2\l(\b_0(9\b_1 \nonumber
\\&+20\b_2+11\b_3)+21\b_1^2+\b_1(63\b_2+30\b_3)\nonumber
\\& +(4\b_2+\b_3)(9\b_2+5\b_3))+\b_1+2\b_2+\b_3)\big]^{-1}\,.
\end{align}

\bibliography{bigravity_refs}{}

\end{document}